\documentclass[a4paper]{spie}  

\usepackage{amsmath,amsfonts,amssymb}
\usepackage{graphicx,xspace}
\usepackage{lastpage}
\usepackage{bm}
\usepackage{siunitx, verbatim}
\usepackage[colorlinks=true, allcolors=blue]{hyperref}

\usepackage[dvipsnames]{xcolor} 
\usepackage{xparse} 
\usepackage[normalem]{ulem} 
\usepackage{soulutf8} 

\newcommand{\revA}[1]{{#1}}

\newcommand{\eq}[1]{Eq.\,({#1})}
\newcommand{\cf}{\emph{cf.}~}

\title{Extending AMIRAL's blind deconvolution of Adaptive Optics corrected images with  Markov chain Monte Carlo methods}

\author[a]{Alix Yan}
\author[a]{Laurent M.\ Mugnier}
\author[b]{Jean-François Giovannelli}
\author[a]{Romain Fétick}
\author[a]{Cyril Petit}
\affil[a]{DOTA/ONERA - Université Paris Saclay, 29 Avenue de la Division Leclerc, Châtillon, France} 
\affil[b]{IMS (Univ. Bordeaux, CNRS, BINP), 351 Cours de la libération, Talence, France}

\authorinfo{Further author information: (Send correspondence to Alix Yan)\\Alix Yan: E-mail: alix.yan@onera.fr, Telephone: +33 (0)1 46 73 47 89  }

\begin{document} 
\maketitle

\begin{abstract}
Adaptive optics (AO) corrected image restoration is particularly difficult, \revA{as it suffers}  from the lack of knowledge on the point spread function (PSF) in addition to usual difficulties. 
An efficient approach is to marginalize the object out of the problem and to estimate the PSF and (object and noise) hyperparameters only, before deconvolving the object using these \revA{estimates}. 
Recent works have applied this marginal blind deconvolution method, combined to a parametric model of the PSF, to a series of AO corrected astronomical and satellite images. 
\revA{In this communication, we propose a new restoration method, which consists in choosing the Minimum Mean Square Error (MMSE) estimator and computing the latter thanks to a Markov chain Monte Carlo (MCMC) algorithm.}
\revA{We validate our method by means of realistic simulations, in two very different contexts:  an astronomical observation on VLT/SPHERE and a ground-based LEO satellite observation on a 1.52\,m telescope. Finally, we present results on experimental images for both applications.}

\end{abstract}


\section{INTRODUCTION}
\label{sec:intro}  

Ground-based high angular resolution imaging in the visible has numerous applications, such as astronomy and satellite observation. These observations are limited by atmospheric turbulence, which can be corrected in real time by adaptive optics (AO). However, the correction is partial and residual blurring remains, impacting high spatial frequencies of the observed object. Therefore, the observation system includes post-processing to restore the high frequencies~\cite{Idier08a}.

The residual blurring is described by the system point-spread function (PSF) which is not entirely known, so both the observed object and the PSF are estimated. The historical way to proceed was to estimate them jointly~\cite{Mistral}, leading to a degenerated solution~\cite{Blanco}. Another way to proceed is to first estimate the PSF by ``marginalizing” over the object, \textit{i.e.} by integrating the probability density function over all possible objects in a given class, and then deconvolve the image with respect to the estimated PSF~\cite{Blanco,Levin}. In our case, the PSF has a physical parametric model and the object is described by a Gaussian prior with a parametric model for its power spectral density (PSD), whose parameters are also estimated along with PSF parameters. The method we have been using \revA{so far}, AMIRAL (standing for Automatic Myopic Image Restoration ALgorithm), combines PSF and PSD parametrization as well as a marginal maximum \textit{a posteriori} (MAP) estimator \cite{MNRAS2020}.

\revA{The present method uses another Bayesian estimator, which minimizes the mean square error on the sought parameters. This enables us to include uncertainties on PSF and PSD parameters~\cite{Orieux2,Villeneuve}.} 
To do so, we include prior distributions for PSF and PSD parameters and we approximate the full posterior distribution by stochastic sampling. 
We first introduce \revA{a} Markov chain Monte Carlo (MCMC) \revA{algorithm} to sample the posterior distribution~\cite{Orieux1}. 
Using the drawn samples, we compute the posterior mean, \textit{i.e.} the minimum mean square error (MMSE) estimator, and the posterior standard deviations.
\revA{Then, we validate our method on simulated data and we finally apply our method to experimental data, for both astronomical and satellite observation contexts, corresponding to two very different instruments and turbulence conditions.} 

\section{Imaging model and marginal MMSE estimator}

\subsection{Imaging equation and prior distributions}
We consider that the image $\bm{i}$ results from the 2D discrete convolution of the object $\bm{o}$ with the PSF $\bm{h}$, to which noise $\bm{n}$ (mostly photon noise and detector readout noise) is added, giving the following imaging model~\cite{Demoment}:
\begin{align}
\bm{i} = \bm{h}*\bm{o}+\bm{n}
\label{imform}
\end{align}
\revA{where $\bm{i}$, $\bm{o}$, $\bm{h}$ and $\bm{n}$ are arrays of size $N \times N$, with $P = N^2$ pixels.}

For this study, we simulate \revA{the image restoration in two different contexts: for astronomical images on the VLT/SPHERE with AO system SAXO and Zimpol imaging polarimeter~\cite{MNRAS2020}, and for ground-based satellite observation at the Côte d'Azur Observatory (OCA) with AO system ODISSEE~\cite{petit-p-20}.} Noise is \revA{taken independent from the object, and is} approximated as zero-mean, additive, white and Gaussian, with an inverse variance of $\gamma_n$. Therefore its PSD is $S_n = 1/\gamma_n$. 

\revA{For the PSF, we use the PSFAO19 model~\cite{AA2019}, which has been designed specifically for describing an AO-corrected PSF with few physical parameters. Its two main parameters are the Fried parameter $r_0$ taken at the imaging wavelength (850\,n\si{\meter}), describing the turbulence's strength, and the variance of the residual turbulent phase $v_{\phi}$, describing the quality of AO correction. This model has been validated on several AO systems and on very different telescopes~\cite{AA2019,petit-p-20}.}

\begin{figure}[htb]

\begin{minipage}[b]{1.0\linewidth}
  \centering
  \centerline{\includegraphics[width=0.6\linewidth]{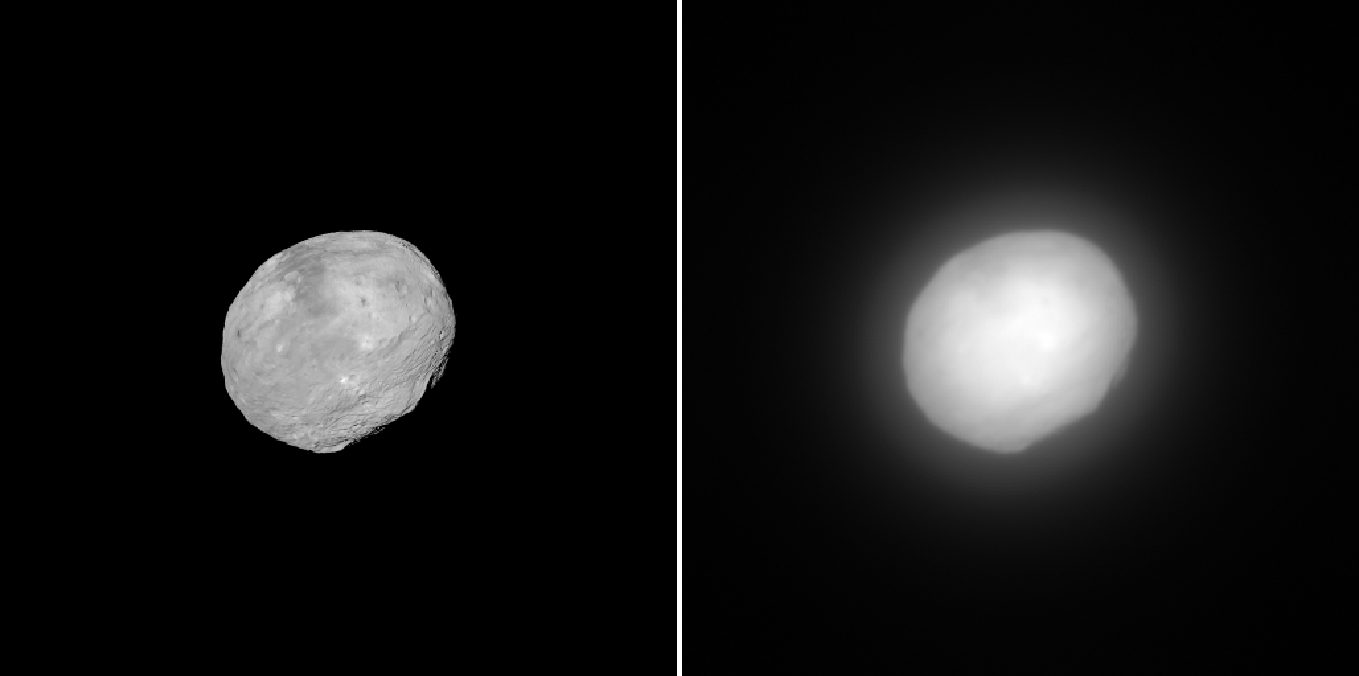}}
\end{minipage}
\caption{Left: true object $\bm{o}$ (synthetic view of Vesta), \revA{of size $512 \times 512$}. Right: simulated image $\bm{i}$ \revA{of size $512 \times 512$}, with true parameters $r_0 = 0.15$\,\si{\meter}, $v_{\phi} = 1.3$\,\si{\radian\squared} and $\gamma_n = 2.62$ $10^{-4}$\,ph$^{-2}$.}
\label{fig:obj}
\end{figure}

As a prior for the object, we consider a Gaussian model described by its mean $\bm{m}_o$ and its PSD $\bm{S}_o$. \revA{The mean object $\bm{m}_o$ is taken uniform on all pixels, equal to the mean value of the image (because the uncertainty on $\bm{m}_o$ is small given the number of pixels). For the object PSD, we use the following parametric model:}
$$S_o(\bm{f}) = \dfrac{1}{\gamma_o}\bar{S_o}(\bm{f}) \text{, with        } \bar{S_o}(\bm{f}) = 1/(k+f^p)$$ 
and $f=|\bm{f}|$ the radial frequency. This circularly-symmetric model is a slightly modified writing of Matérn's model~\cite{Conan98,RamaniVBU08}.
In this model, $\gamma_o$ sets the \revA{global} PSD level, $p$ describes the PSD decrease at high frequencies, and $k$ gives the breakpoint between the two regimes \revA{of the model}. \revA{In previous works~\cite{MNRAS2020}, attempts to estimate hyperparameter $p$ jointly with the other sought parameters} has been shown to strongly decrease PSF parameters estimation accuracy. Therefore, we choose to work in a ``mostly supervised" mode, where $p$ is fixed to a standard value. \revA{In the case of astronomical observations of asteroids, a well-fitting empirical value is around $p=3$, whereas for satellite observation a standard value for $p$ would be around 2.5 - 2.6.}

\revA{By making a circulant approximation for the object covariance matrix,} we can write the object prior distribution as follows:
$$
    p(\bm{o}|\gamma_o,k) = \Big(\dfrac{\gamma_o}{2\pi}\Big)^{P/2} \prod_{\bm{f}} \Big( \bar{S_o}(\bm{f})^{-1/2}
    \times \exp \big[-\dfrac{\gamma_o}{2}\bar{S_o}(\bm{f})^{-1}|\Tilde{o}(\bm{f})-\Tilde{m}_{o}(\bm{f})|^2 \big] \Big)
$$
where $\Tilde{.}$ denotes the discrete Fourier transform (DFT) and the product on $\bm{f}$ is on all pixels in the spatial frequency domain. For the object and the image, the DFT is normalized so as to comply with Parseval's theorem. For the PSF, the DFT is normalized so that $\Tilde{h}(0)$ equals the sum of the PSF. Moreover, this value is normalized to 1.

\revA{Given the hypothesis made on the noise model, and the imaging model (\cf    \eq{\ref{imform}}) which we approximate as periodized,} the likelihood writes:
$$
    p(\bm{i}|\bm{o},\gamma_n,r_0,v_{\phi})
    = \Big(\dfrac{\gamma_n}{2\pi}\Big)^{P/2} \prod_{\bm{f}} \exp(-\dfrac{\gamma_n}{2}|\Tilde{i}(\bm{f})-\Tilde{h}(\bm{f})\Tilde{o}(\bm{f})|^2)
$$
where $\Tilde{h}$ is the (discretized) optical transfer function (OTF).

Regarding PSF parameters as well as noise and object PSD parameters, hereafter called parameters, we consider that each parameter $\gamma_n$, $\gamma_o$, $k$, $r_0$ and $v_{\phi}$ can take any value in a given range. Therefore, following the Laplace rule (or principle of insufficient reason), we use uniform priors for each of them~\cite{Kass}. Additionally, all parameters are considered independent. 

Image $\bm{i}$ and object PSD parameters $\gamma_o$ and $k$ are chosen independent conditionally to the other parameters, meaning $p(\bm{i},\gamma_o,k|\bm{o},\gamma_n,r_0,v_{\phi}) = p(\bm{i}|\bm{o},\gamma_n,r_0,v_{\phi}) \times p(\gamma_o,k|\bm{o},\gamma_n,r_0,v_{\phi})$ therefore $p(\bm{i}|\bm{o},\gamma_n,\gamma_o,k,r_0,v_{\phi}) = p(\bm{i}|\bm{o},\gamma_n,r_0,v_{\phi})$. 
Finally, the object, the noise variance and the PSF parameters are taken independent conditionally to object PSD parameters. 

The joint distribution is, following the \revA{conditioning rule}, the multiplication of the likelihood by the prior distributions: 
$$
    p(\bm{i},\bm{o},\gamma_n,\gamma_o, k, r_0,v_{\phi}) = p(\bm{i}|\bm{o},\gamma_n,r_0,v_{\phi}) 
    \times p(\bm{o}|\gamma_o,k)p(\gamma_n)p(\gamma_o)p(k)p(r_0)p(v_{\phi})
$$
Following the Bayes rule, the posterior distribution is:
$$p(\bm{o},\gamma_n,\gamma_o,k,r_0,v_{\phi}|\bm{i})=\dfrac{p(\bm{i},\bm{o},\gamma_n,\gamma_o,k,r_0,v_{\phi})}{p(\bm{i})}$$
with $p(\bm{i})$ a constant with respect to the unknowns.
\subsection{Marginal estimator}
As explained above, a way to estimate the parameters and the object is to first estimate the parameters by marginalizing the posterior probability density over the object. We can also write the marginal distribution using Bayes rule:
\begin{align}
p(\gamma_n,\gamma_o,k,r_0,v_{\phi}|\bm{i}) &= \int p(\bm{o},\gamma_n,\gamma_o,k,r_0,v_{\phi}|\bm{i}) \mathrm{d} \bm{o} \nonumber \\
&= \dfrac{p(\gamma_n)p(\gamma_o)p(k)p(r_0)p(v_{\phi})}{p(\bm{i})} \times p(\bm{i}|\gamma_n, \gamma_o, k, r_0, v_{\phi}) \label{margpost}
\end{align}
The prior distributions for $\gamma_n$, $\gamma_o$, $k$, $r_0$ and $v_{\phi}$ were given above. Given that the noise and the object are considered Gaussian, the image being a linear combination of both is also Gaussian. Therefore, the marginal likelihood writes\footnote{\revA{Given \eq{\ref{marglike}}, maximizing the marginal likelihood, as in the previous method~\cite{Blanco,MNRAS2020}, can be interpreted as finding the parameters for our image PSD model $S_i$ (\cf\eq{\ref{imagepsd}}) that best fit the empirical PSD $|\Tilde{i}(\bm{f})-\Tilde{m}_{i}(\bm{f})|^2$.}}:
\begin{align}
    p(\bm{i}|\gamma_n,\gamma_o,& k,r_0,v_{\phi}) = (2\pi)^{-P/2} \prod_{\bm{f}} \Big(S_i(\bm{f})^{-1/2} \times \exp\big[-\dfrac{1}{2}|\Tilde{i}(\bm{f})-\Tilde{m}_{i}(\bm{f})|^2/S_i(\bm{f}) \big] \Big) \label{marglike}
\end{align}
with image PSD $\bm{S}_i$ and mean image $\bm{m}_i$:
\begin{align}
S_i(\bm{f}) &= S_o(\bm{f})|\Tilde{h}(\bm{f})|^2+S_n \label{imagepsd}\\
\Tilde{m}_{i}(\bm{f}) &= \Tilde{h}(\bm{f})\Tilde{m}_{o}(\bm{f}) \nonumber
\end{align}
The marginal posterior \revA{distribution} for the parameters can then easily be written using \eq{\ref{margpost}} - \eq{\ref{marglike}}.

\subsection{MMSE estimator and sampling}

Given the complexity of the posterior mean, there is no known analytical way to calculate it. 
A way to approximate it is to draw samples under the posterior distribution and calculate their empirical mean. \revA{The posterior distribution being complex, it is not possible to sample it directly, therefore we use a Metropolis-Hastings algorithm to bypass the problem. It consists in drawing samples under a chosen proposition distribution and accepting the samples (else, duplicating the previous value) with a given probability. Several versions are possible, in particular: we can either draw all the parameters simultaneously (standard Metropolis-Hastings) or separately (Metropolis-Hastings-within-Gibbs). Drawing the parameters together 
can make the acceptance probability fall (except if we use more advanced gradient-based algorithms such as MALA or HMC methods), whereas drawing parameters individually requires more likelihood computations, which can slow down the algorithm.} 

\revA{For simplicity, we use here the second version. In a standard Gibbs algorithm, each parameter is drawn under its own conditional posterior distribution, which is proportional to the prior of the considered parameter times the marginal likelihood (\cf\eq{\ref{marglike}}). As mentioned above, because drawing the parameters under their conditional posterior distribution is difficult, we use a Metropolis-Hastings-within-Gibbs algorithm instead of the standard Gibbs. Asymptotically, the samples are under the posterior distribution for all unknowns, and their empirical mean tends towards the expectation of the distribution.} 

\revA{In our case, we use a Random Walk Metropolis-Hastings algorithm: here, the proposed sample for each parameter is drawn under a symmetric (Gaussian) distribution around the current value of the parameter. The standard deviation of the Gaussian distribution is chosen to be $0.01$ times the allowed range of the prior. This choice is based on the physical meaning of the parameters, according to the sensitivity of the PSF or the noise or object PSD to the parameter, and only impacts the convergence time.}

\revA{(Note that once we have drawn the samples, we could also compute the posterior median, which is another optimal estimator, namely the one minimizing the mean absolute error.)}

\revA{\section{Results on astronomical images}}
\revA{\subsection{Results on simulated data}}
The obtained results are shown for the simulated image described above, using as the true object the synthetic view of asteroid Vesta\revA{, built by the OASIS software~\cite{OASIS}, on a dark background of size $512 \times 512$ pixels. True PSF parameters are $r_0 = 0.15$\,\si{\meter} and $v_{\phi} = 1.3$\,\si{\radian\squared}. The AO system parameters are taken identical to those used with the previous method~\cite{MNRAS2020}, for comparison purposes.} Noise is taken zero-mean, additive, white and Gaussian with a variance equal to the mean value of the object as a first approximation of the photon noise. The total flux of the object is set to $F_{o} = 10^9$\,ph (photons), typical from VLT/SPHERE/Zimpol asteroid observations (ESO Large Program ID 199.C-0074), therefore $\gamma_n = P / F_{o} = 2.62 \times 10^{-4}\,\text{ph}^{-2}$. 

\begin{figure}[!htb]
\centerline{\includegraphics[width=0.75\linewidth]{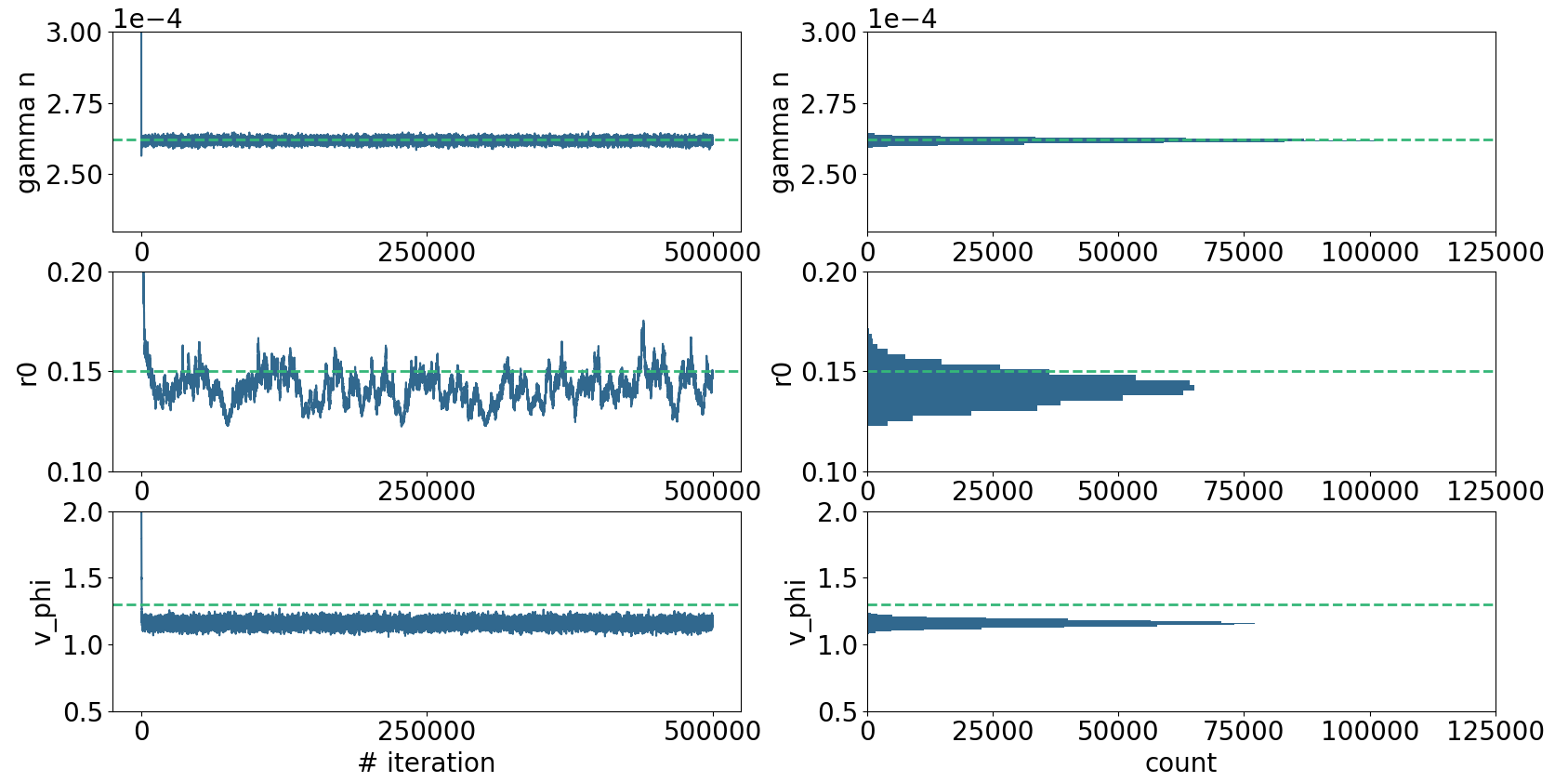}}
\caption{From top to bottom: $\gamma_n$, $r_0$, $v_{\phi}$. Left: \revA{chain of samples for simulated astronomical images}. Right: corresponding histogram. (True values in dashed line.)}
\label{fig:ch_hist}
\end{figure}

The PSF and PSD parameters are estimated following the proposed method, except the object PSD power which is fixed to $p=3$. The Gibbs sampler is run for 500 000 iterations, which corresponds to a few hours, to verify the convergence.
In Figure \ref{fig:ch_hist}, we plot the samples chains and the corresponding histograms for $\gamma_n$, $r_0$ and $v_{\phi}$. Their true value is also represented.

The inspection of Figure~\ref{fig:ch_hist} suggests that chains have a short burn-in period, followed by a stationary state. As expected, the samples are correlated to their neighbors, the chains have a Markov feature (particularly noticeable on $r_0$ chain).  
Moreover, the samples are concentrated in a small interval relatively to their mean value.

\begin{table}[!htbp]
    \centering
      \begin{tabular}{ | l | c | c | }
         \hline
         Parameter & $m \pm \sigma$ & True \\ \hline
         $\gamma_n$ ($\text{ph}^{-2}$) & 2.62$\times10^{-4}$ $\pm$ 2.93$\times10^{-5}$ & 2.62$\times10^{-4}$ \\ \hline
         $\gamma_o$ ($\text{ph}^{-2}$) & 2.35$\times10^{-13}$ $\pm$ 9.10$\times10^{-14}$ & - \\ \hline
         $k$ & 0.773 $\pm$ 0.592 & - \\ \hline
         $r_0$ (\si{\meter}) & 0.142 $\pm$ 0.011 & 0.15 \\ \hline
         $v_{\phi} (\si{\radian\squared})$ & 1.16 $\pm$ 0.04 & 1.30 \\ \hline
       \end{tabular}
    \caption{Mean value, associated standard deviation and true value, for $\gamma_n$, $\gamma_o$, $k$, $r_0$ and $v_{\phi}$.}
    \label{tab1}
\end{table}

The empirical mean values $m$ and standard deviation $\sigma$ for each parameter are displayed in Table \ref{tab1}. \revA{The} noise inverse variance is accurately estimated, with an error smaller than 0.2\%. 
The standard deviation for $\gamma_n$, relatively to its mean value, is lower than the one for $\gamma_o$. This latter fact can be explained by the fact that noise is directly observed whereas the object is not (the object is convolved by the PSF). As a consequence, $\gamma_n$ is more reliably estimated than $\gamma_o$. 
Concerning the PSF parameters estimation, \revA{we notice that} the true value for parameter $r_0$ is in the range $m_{r_0} \pm \sigma_{r_0}$ (5\% error), and \revA{the} true $v_{\phi}$ is in the interval $m_{v_{\phi}} \pm 4\sigma_{v_{\phi}}$ (11\% error). \revA{The mean values of $r_0$ and $v_{\phi}$} are very close to the previous results obtained with AMIRAL: for the same conditions, the estimated PSF parameters were $r_0 = 0.142$\,\si{\meter} and $v_{\phi} = 1.13$\,\si{\radian\squared}~\cite{MNRAS2020}, to be compared to the values obtained with the new method (Table~\ref{tab1}) which includes uncertainties.

\begin{figure}[!htb]
\centerline{\includegraphics[width=0.68\linewidth]{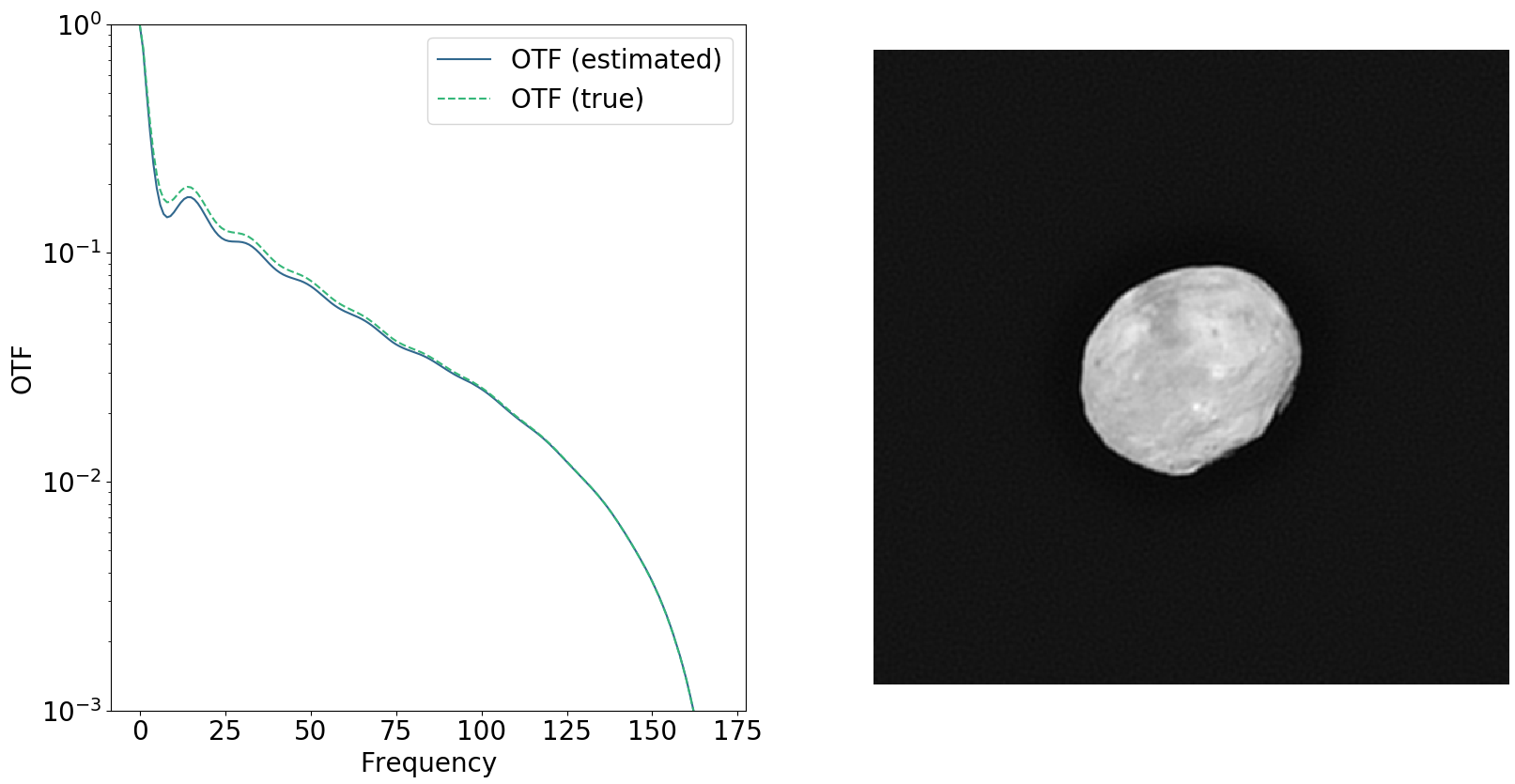}}
\caption{Left: estimated and true OTF (true in dashed line), optical cut-off at pixel 171. Right: restored object (Wiener filtering) \revA{from image in Figure~\ref{fig:obj}, right, of size $512 \times 512$}.}
\label{fig:otfres}
\end{figure}

We also compare the resulting OTF to the true OTF in Figure~\ref{fig:otfres}. The slight underestimation of $r_0$ leads to the lowering of the global OTF level and its impact can mainly be seen at low frequencies. Concerning $v_{\phi}$, its mild underestimation leads to a slower decrease of the OTF and impacts the slope of the latter at medium-high frequencies~\cite{MNRAS2020}. Thus, we notice that the errors on both parameters partially compensate, as a result the normalized RMSE for the OTF is quite small (around 7\%).

\begin{figure}[!htb]
  \centerline{\includegraphics[width=0.68\linewidth]{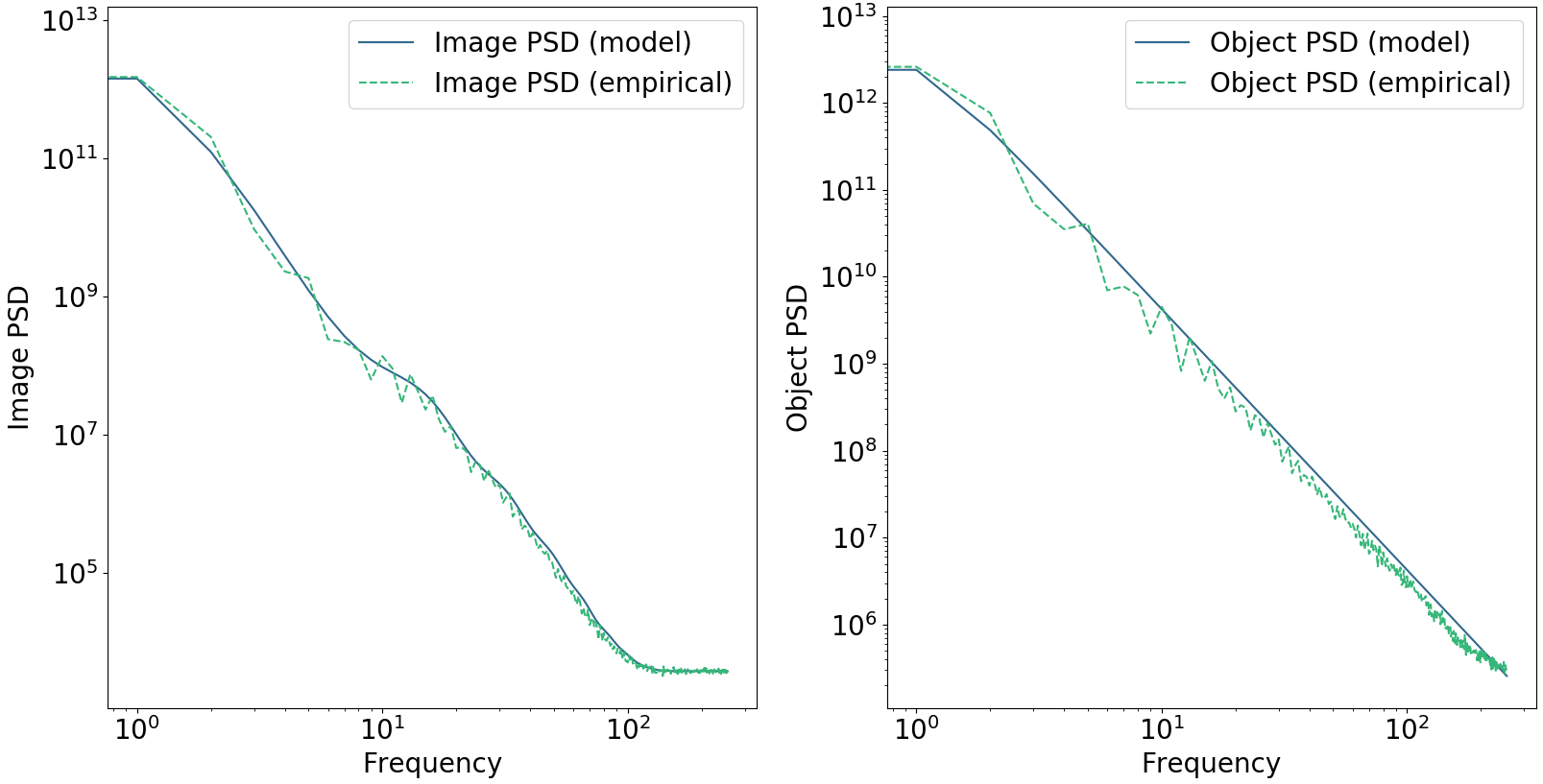}}
\caption{\revA{Vesta simulation PSDs. Model in solid line, and empirical PSD averaged azimuthally in dashed line. Left: Image PSD. Right: Object PSD.}}
\label{fig:psds}
\end{figure}

\revA{An important sanity check of the method is to verify that our model for the image PSD of \eq{\ref{imagepsd}}, which combines object PSD, PSF and noise PSD, accurately fits the empirical image PSD averaged azimuthally (\cf\eq{\ref{marglike}})\revA{. We do so} in Figure~\ref{fig:psds} (left). 
Moreover, given the fact that the true object is not the realization of a Gaussian random field following our PSD model, a way to check $\gamma_o$ and $k$\revA{'s} estimation accuracy is to look at the fitting of our model to the empirical object PSD, averaged azimuthally. As displayed in Figure~\ref{fig:psds} (right), the object PSD model visually fits correctly the empirical object PSD, the slight overestimation of the object PSD being consistent with the slight underestimation of the OTF.} 

\revA{Finally,} Figure~\ref{fig:otfres} shows the image in Figure~\ref{fig:obj} restored with the estimated OTF. Many details of the Vesta surface can be seen, that were not visible on the data.

\begin{figure}[htb]
\centerline{\includegraphics[width=0.75\linewidth]{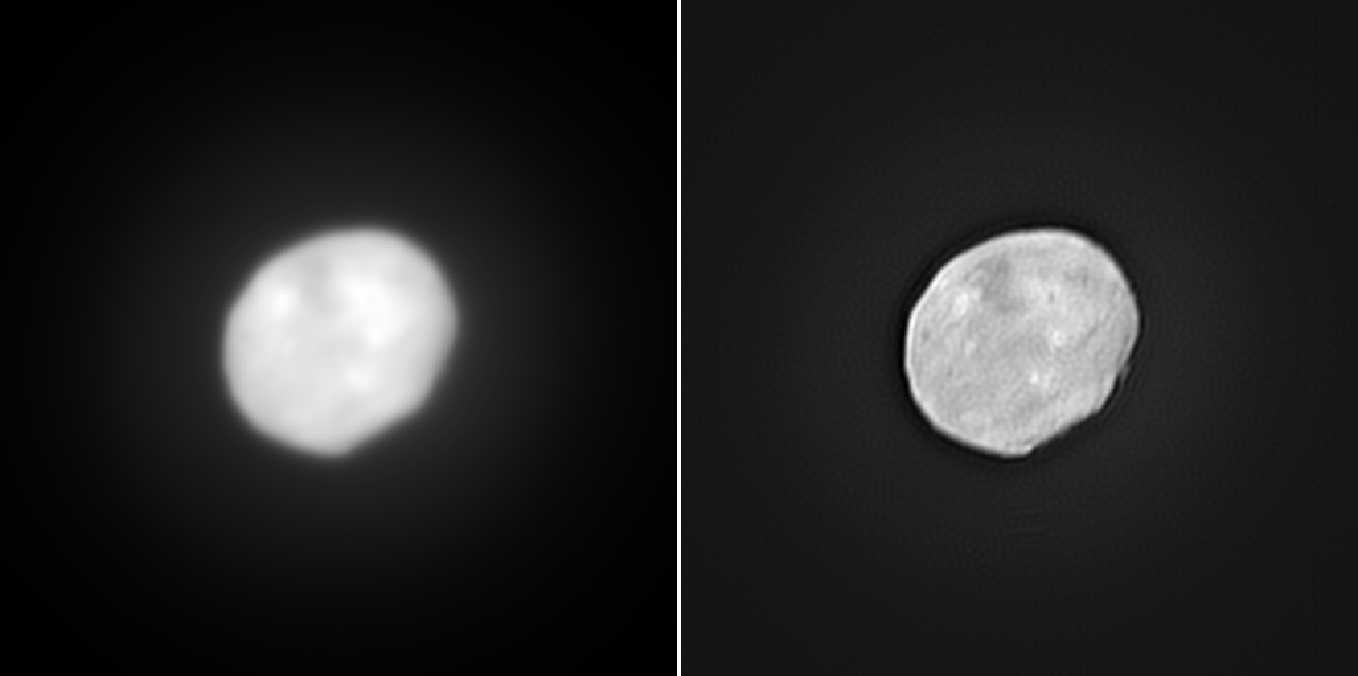}}
\caption{Left: Vesta observed by SPHERE/Zimpol on the European Very Large Telescope (VLT) in Chile~\cite{Vestaexp}. Right: restored object (Wiener filtering) \revA{with the estimated PSF.}}
\label{fig:vestaexpe}
\end{figure}

\revA{\subsection{Results on experimental data}}
\revA{We apply} our method to an experimental image of Vesta~\cite{Vestaexp} \revA{taken by SPHERE/Zimpol} in the same mostly unsupervised mode as previously where $p=3$, and run the Gibbs sampler for 100 000 iterations. Data and restored object are shown in \mbox{Figure}~\ref{fig:vestaexpe}. We recognize some surface details from the synthetic view in Figure~\ref{fig:obj}. 

\begin{figure}[!htb]
  \centerline{\includegraphics[width=0.68\linewidth]{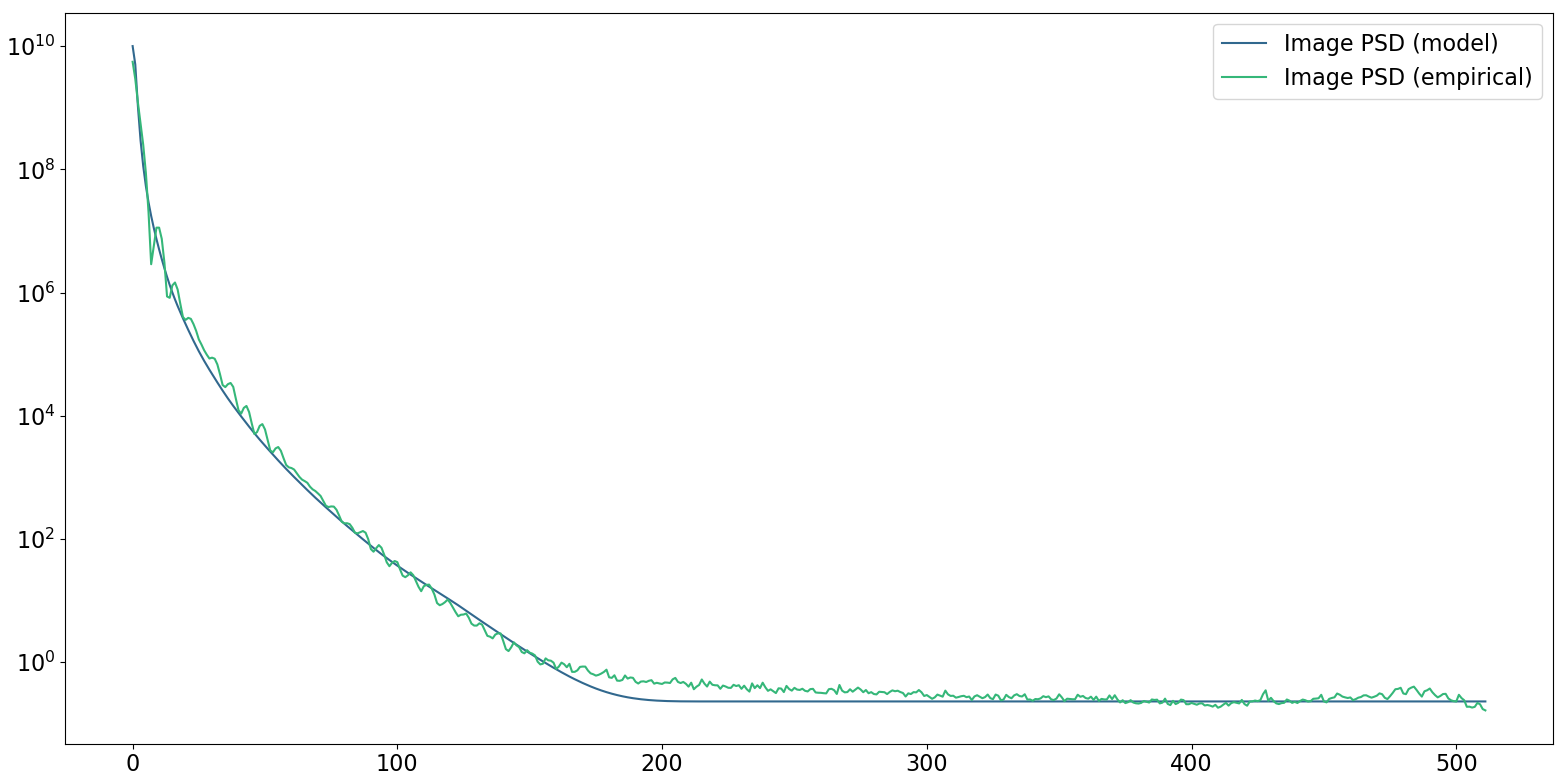}}
\caption{\revA{PSD model and Vesta empirical image PSD averaged azimuthally, in dashed line.}}
\label{fig:vestaexpepsd}
\end{figure}

\revA{Results obtained for the PSF parameters (mean$\pm$(standard deviation)) are the following:} $r_0 = 0.26(\pm 0.04)$\,\si{\meter} and $v_{\phi} = 2.62(\pm 0.06)$\,\si{\radian\squared}.
These values are close to the values obtained with AMIRAL~\cite{MNRAS2020} ($r_0 = 0.32$\,\si{\meter}, $v_{\phi} = 2.78$\,\si{\radian\squared}) for the same conditions\revA{, the newly estimated $r_0$ being more likely than the one estimated by AMIRAL according to the known statistics on $r_0$~\cite{AA2019}. We also look at the image PSD model and the empirical image PSD for Vesta in Figure~\ref{fig:vestaexpepsd}. Both fit well, especially at low \revA{and medium} frequencies\revA{, where signal dominates noise}. For high frequencies, where the noise is dominant, \revA{we see that the noise floor is not flat (whereas we model the noise as white), and believe it may be due to} the data reduction by SPHERE/Zimpol's pipeline.}

\section{Results on satellite images}

\subsection{Results on simulated data}

\revA{We now show results for a simulated satellite image, using as the true object a synthetic view of the SPOT satellite on a dark background of size $512 \times 512$ pixels. We simulate its observation using the ODISSEE AO system at OCA~\cite{petit-p-20}, and with true PSF parameters $r_0 = 0.10$\,\si{\meter} and $v_{\phi} = 1.85$\,\si{\radian\squared}, which corresponds to the stronger turbulence of a non-astronomical site, and to a more modest correction than for the astronomical simulation. 
 Noise is taken zero-mean, additive, white and Gaussian, and its variance is taken equal to the mean value of the object. Here, the mean flux is $10^{4}\,\text{ph/pix}$, corresponding to an optimistic yet realistic value.} 
\begin{figure}[htb]
\begin{minipage}[b]{1.0\linewidth}
  \centering
  \centerline{\includegraphics[width=0.6\linewidth]{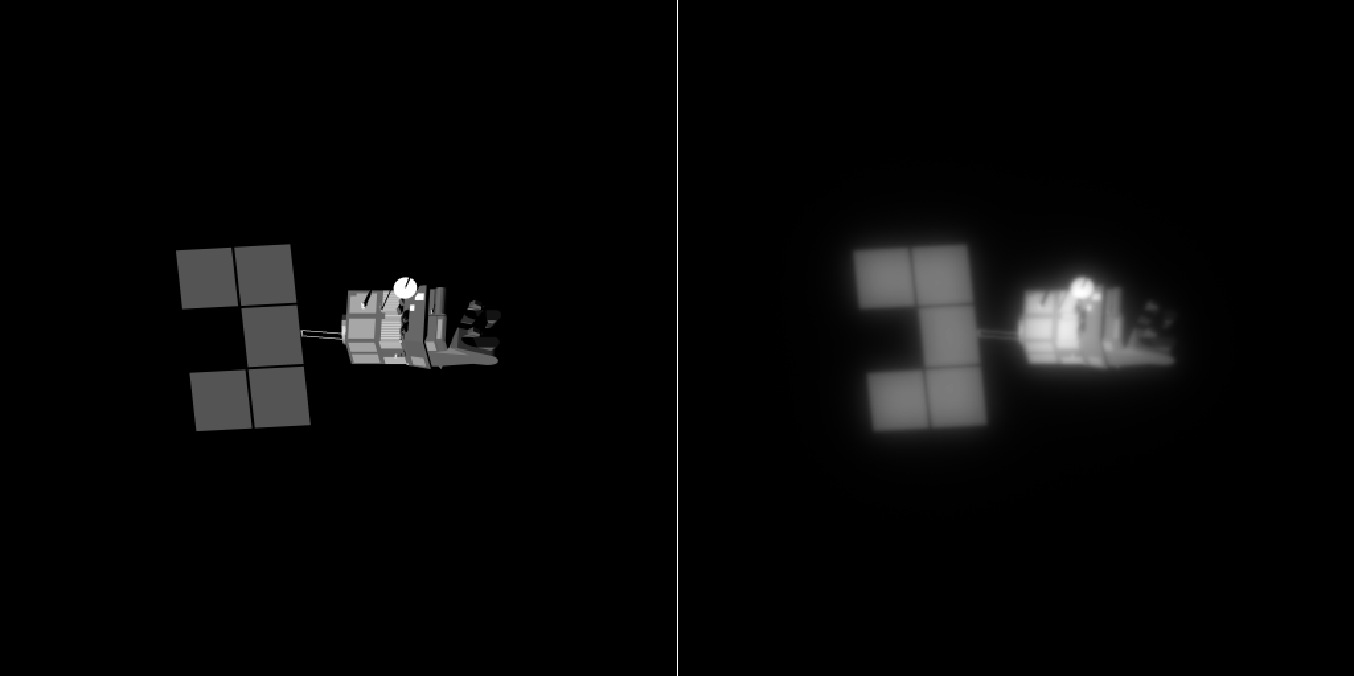}}
\end{minipage}
\caption{\revA{Left: true object (synthetic view of SPOT). Right: simulated image, with true parameters $r_0 = 0.10$\,\si{\meter}, $v_{\phi} = 1.85$\,\si{\radian\squared} and $\gamma_n = 2.62$ $10^{-4}$\,ph$^{-2}$.}}
\label{fig:spotobj}
\end{figure}

\begin{figure}[htb]
\centerline{\includegraphics[width=0.75\linewidth]{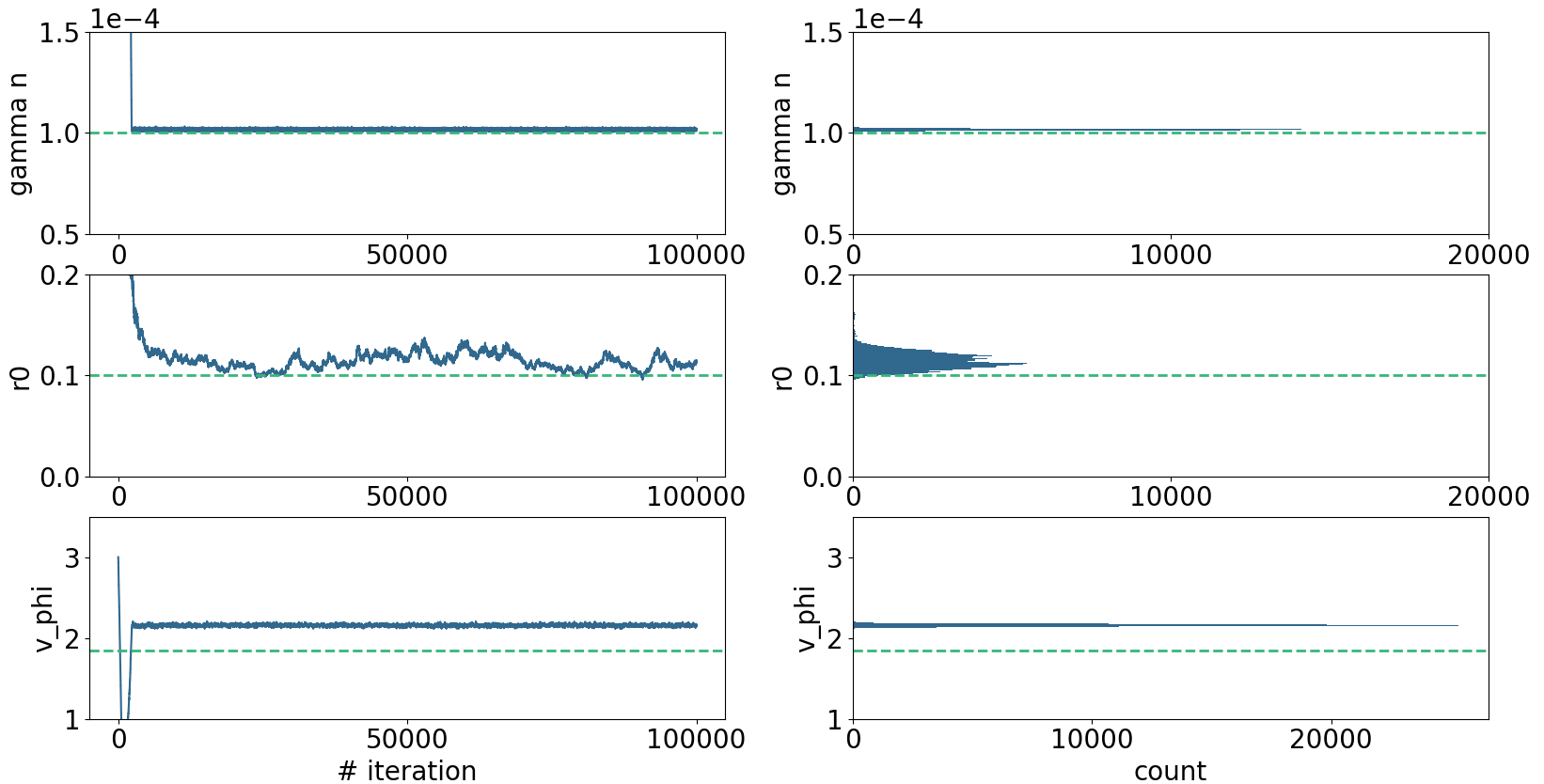}}
\caption{From top to bottom: $\gamma_n$, $r_0$, $v_{\phi}$. Left: \revA{chains of samples for simulated satellite images}. Right: corresponding histogram. (True values in dashed line.)}
\label{fig:chainsat}
\end{figure}

\revA{The object PSD power $p$ is fixed to an empirical standard value for satellites $p=2.6$, to fit the empirical object PSD. 
The Gibbs sampler is run for 100 000 iterations. In Figure~\ref{fig:chainsat}, we plot the samples chains and the corresponding histograms for $\gamma_n$, $r_0$ and $v_{\phi}$. Their true value is also represented.}

\begin{table}[htbp]
    \centering
      \begin{tabular}{ | l | c | c | }
         \hline
         Parameter & $m \pm \sigma$ & True \\ \hline
         $\gamma_n$ ($\text{ph}^{-2}$) & 1.02 $\times10^{-4}$ $\pm$ 3.63$\times10^{-7}$ & 1.00 $\times10^{-4}$ \\ \hline
         $\gamma_o$ ($\text{ph}^{-2}$) & 1.90 $\times10^{-14}$ $\pm$ 2.37 $\times10^{-15}$ & - \\ \hline
         $k$ & 3.14 $\pm$ 1.45 & - \\ \hline
         $r_0$ (\si{\meter}) & 0.115 $\pm$ 0.008 & 0.10 \\ \hline
         $v_{\phi} (\si{\radian\squared})$ & 2.16 $\pm$ 0.01 & 1.85 \\ \hline
       \end{tabular}
    \caption{\revA{Mean value, associated standard deviation and true value, for $\gamma_n$, $\gamma_o$, $k$, $r_0$ and $v_{\phi}$.}}
    \label{tab2}
\end{table}

\revA{The empirical mean values $m$ and standard deviation $\sigma$ of the posterior distribution for each parameter are displayed in Table \ref{tab2}. The noise inverse variance $\gamma_n$ as well as PSF parameters $r_0$ and $v_{\phi}$ are quite well estimated, with an error of respectively 2\%, 15\% and 17\%. These results are quite close to those obtained with AMIRAL: for similar conditions, the estimated PSF parameters are $r_0 = 0.107$\,\si{\meter} and $v_{\phi} = 2.09$\,\si{\radian\squared}. Concerning the posterior standard deviation: for both PSF parameters, the estimated $\sigma$ seems small, giving an under-estimated uncertainty. By complementary simulations, we have checked that the difficult estimation of PSF parameters, and so the underestimation of uncertainties, is due to the strong turbulence conditions and modest AO correction.} 


\begin{figure}[htb]
\centerline{\includegraphics[width=0.68\linewidth]{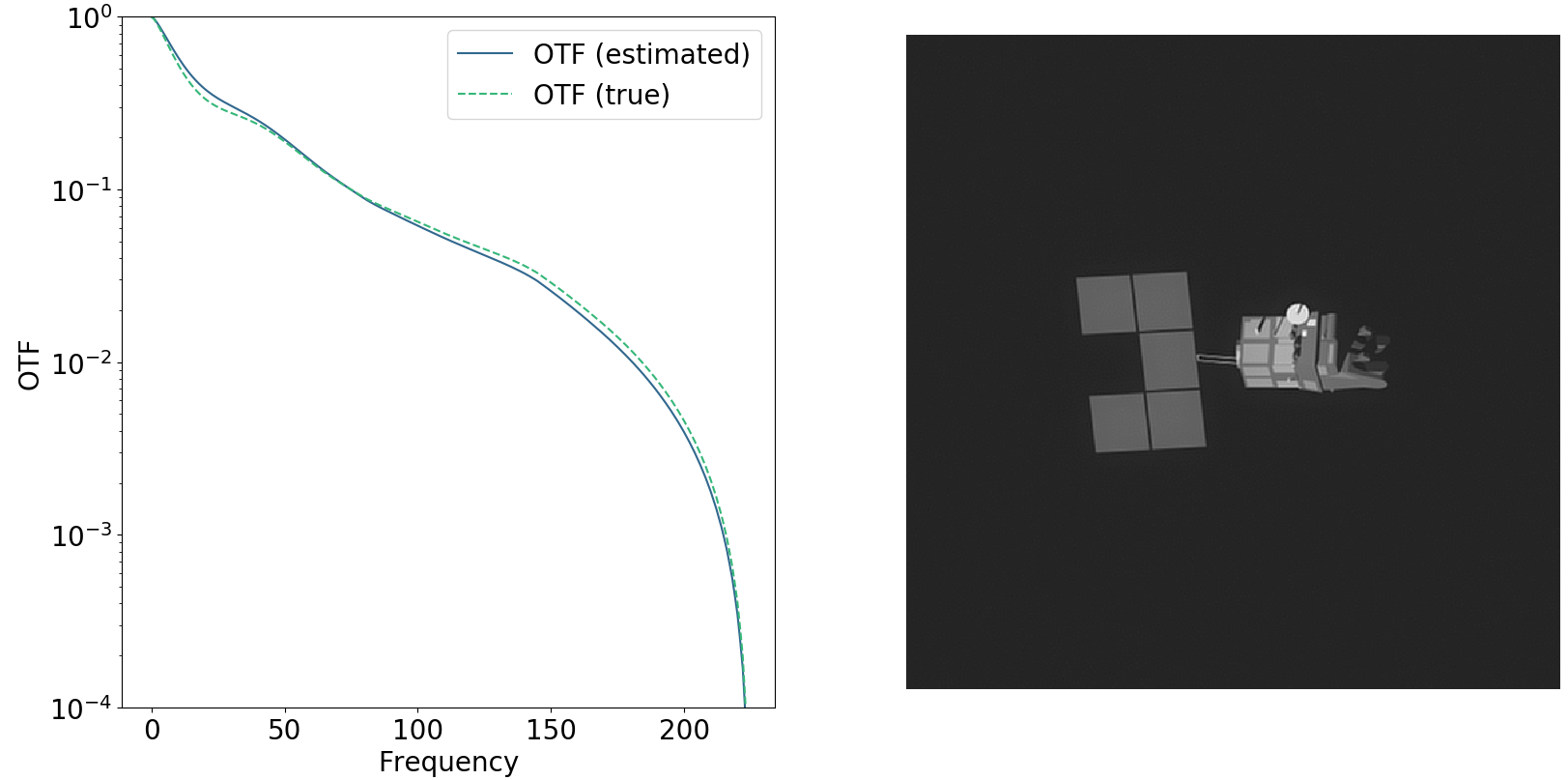}}
\caption{Left: estimated and true OTF (true in dashed line). Right: restored object (Wiener filtering) \revA{from simulated image in Figure~\ref{fig:spotobj}, right}.}
\label{fig:spototf}
\end{figure}

\begin{figure}[!htb]
  \centerline{\includegraphics[width=0.68\linewidth]{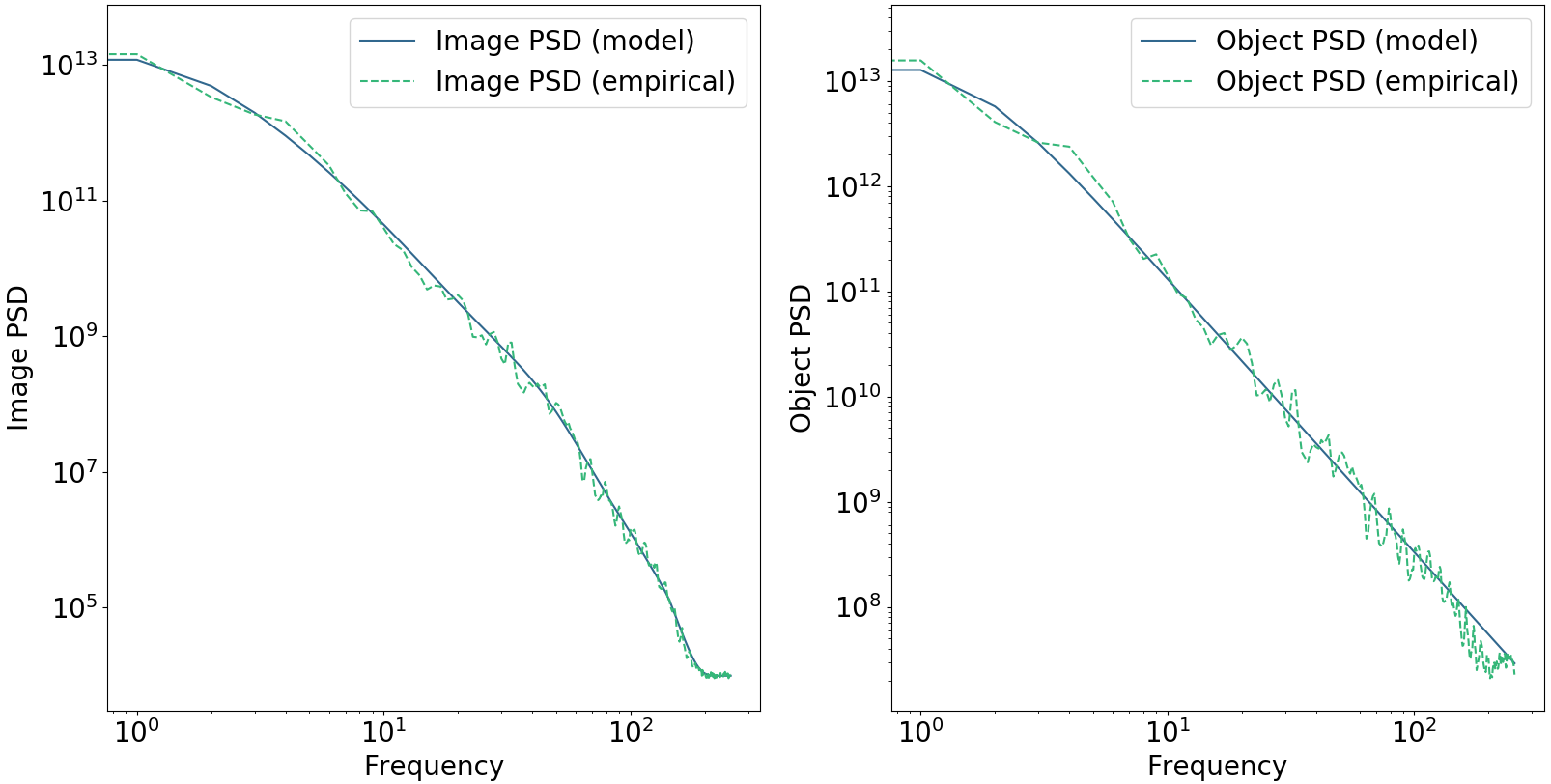}}
\caption{\revA{SPOT simulation PSDs. Model in solid line, and empirical PSD averaged azimuthally in dashed line. Left: Image PSD. Right: Object PSD.}}
\label{fig:spotpsd}
\end{figure}
\newpage
\revA{We also compare the resulting estimated OTF to the true OTF in Figure~\ref{fig:spototf}. Here again, as discussed previously, we notice that the errors on both parameters partially compensate, as a result the normalized RMSE for the OTF is quite low (around 8\%).
\revA{Additionally, the estimations result in a good image PSD fitting as shown in Figure~\ref{fig:spotpsd}. Moreover, as displayed in Figure~\ref{fig:spotpsd}, the object PSD model visually fits well the empirical object PSD.}
Finally, Figure~\ref{fig:spototf} results from the restoration of the image in Figure~\ref{fig:spotobj} (right) using the estimated OTF. We notice that some details of the satellite surface were restored.}

\revA{\subsection{Results on experimental data}}

\revA{Finally, we test our method on an experimental image of Envisat, taken at the OCA with ONERA's ODISSEE AO system~\cite{petit-p-20}. We fix paramater $p$ to a reasonable value for satellites ($p = 2.5$), and run the Gibbs sampler for 500 000 iterations.}

\begin{figure}[htb]
\centerline{\includegraphics[width=0.75\linewidth]{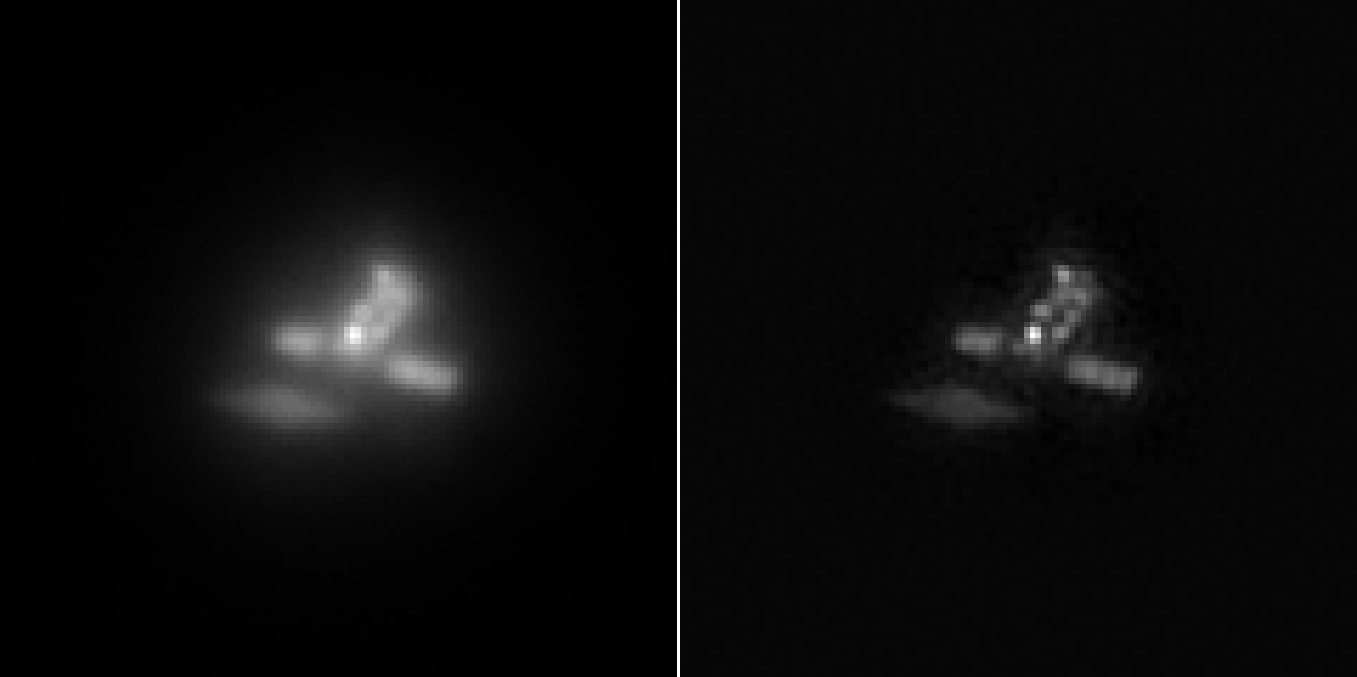}}
\caption{Left: Envisat observed by ODISSEE at the OCA~\cite{petit-p-20}. Right: restored object (Wiener filtering).}
\label{fig:envisatexpe}
\end{figure}

\revA{Our first results on the estimated PSF parameters are the following: $r_0 = 0.08$\,\si{\meter} and $v_{\phi} = 0.89$\,\si{\radian\squared}, which can be compared to the results obtained with AMIRAL: $r_0 = 0.06$\,\si{\meter} and $v_{\phi} = 1.13$\,\si{\radian\squared}, for the same conditions. Concerning the restored image, as shown in Figure~\ref{fig:envisatexpe}, we retrieve some elements of the satellite, and we checked on a CAD model of Envisat that the bright spots we obtain on the restored image indeed correspond to instruments and antennas on its surface.}

\begin{figure}[!htb]
  \centerline{\includegraphics[width=0.68\linewidth]{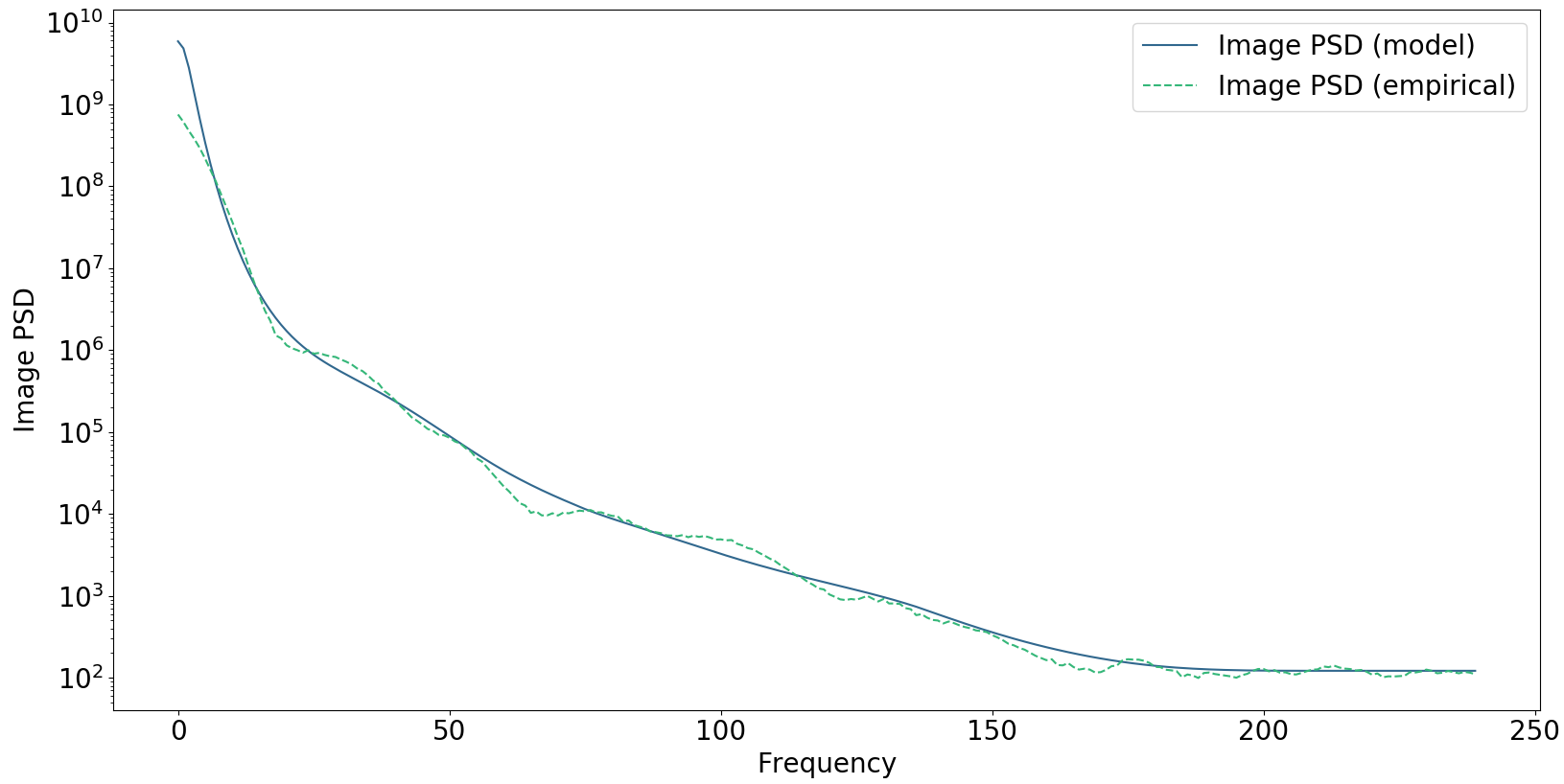}}
\caption{\revA{PSD model and Envisat empirical image PSD averaged azimuthally, in dashed line.}}
\label{fig:envisatexpepsd}
\end{figure}

\revA{Finally, concerning the image PSD model and the empirical image PSD for Envisat, as shown in Figure~\ref{fig:envisatexpepsd}, we have a globally good image PSD fitting. The oscillations of the empirical image PSD are likely to come from oscillations of the OTF, which are consistent with the relatively short exposure time ($\approx1\,$s) and constitute a deviation to the infinite exposure assumption of our AO PSF model. 
These oscillations might also come partly from the spectrum of the object itself, which deserves further studies by means of simulations.}

\section{Conclusion}
\revA{We have presented a new marginal blind deconvolution method extending previous works: in addition to PSF and hyperparameter estimation combined with image restoration, we now have access to the whole posterior distribution. This enables us to compute the optimal estimator minimizing the mean square error. 
Additionally, with the posterior distribution, we can compute uncertainties based on the posterior standard deviation. This new method 
has been validated on a simulated images, giving accurate estimation of noise and object hyperparameters, as well as \revA{satisfactory} PSF parameters estimation. Two very different contexts were simulated: on the one hand, the observation of asteroid Vesta on the VLT, and on the other hand the observation of the SPOT satellite, with more difficult conditions and a less performing AO system (ONERA's AO bench at OCA's $1.52$\,\si{\meter} telescope). The satisfactory results obtained in both conditions suggests the broad applicability of the method. Finally, our new method has also been applied to experimental images, in both contexts.}

In this work, hyperparameter $p$, which codes for the decrease of the object PSD, has been fixed to a reasonable value \revA{according to the class of the object (either asteroid, or satellite). The PSF estimation quality is sensitive to the choice of $p$, as we verified it by changing its value~\cite{GRETSI2022}. Moreover, jointly estimating $p$ with the other parameters is difficult as mentioned in earlier studies~\cite{MNRAS2020}. 
In the near future, we plan to tackle both the estimation of $p$ and the joint processing of an image series to improve the PSF estimation.} 


\revA{Lastly,} we currently sample each parameter individually, using a Metropolis-Hastings-within-Gibbs algorithm and the convergence speed issue has not been investigated. To accelerate the convergence, a possible development is to use a Metropolis-Hastings algorithm to sample the posterior distribution jointly for all parameters, thus without using the Gibbs sampler, and to use gradient-based methods such as a Metropolis-adjusted Langevin algorithm~\cite{robert,vacar}.

\bigskip
\section{Acknowledgement}
This study has been partly funded by the French Aerospace
Lab (ONERA) in the framework of the SUSA Project.

\clearpage 
\bibliography{report} 
\bibliographystyle{spiebib} 

\end{document}